\documentclass[aps,pre,twocolumn,amssymb,amsfonts,floatfix,superscriptaddress]{revtex4}

\usepackage{graphicx}
\usepackage{dcolumn}
\usepackage{bm}
\usepackage{mathbbol}
\usepackage{bbm}

\usepackage{amsmath,amssymb,amsthm}
\usepackage{amsfonts}
\usepackage{times}
\usepackage{graphicx}
\usepackage{bm}
\usepackage{color}
\usepackage{multi row}

\begin{document}

\title{Many-body localization transition through pairwise correlations}
\author{Jaime L. C. da C. Filho}
 \email{jaime.filho@if.uff.br}
\affiliation{Instituto de F\'{i}sica, Universidade Federal Fluminense, Av. Gal. Milton Tavares de Souza s/n, Gragoat\'{a}, 24210-346 Niter\'{o}i, Rio de Janeiro, Brazil}
\affiliation{Instituto Federal do Par\'{a} - Campus Castanhal, \\BR 316 Km 61, Saudade II, 68740-970, Castanhal, PA, Brazil}
\author{Andreia Saguia}
\email{amen@if.uff.br}
\affiliation{Instituto de F\'{i}sica, Universidade Federal Fluminense, Av. Gal. Milton Tavares de Souza s/n, Gragoat\'{a}, 24210-346 Niter\'{o}i, Rio de Janeiro, Brazil}
\author{Lea F. Santos}
\email{lsantos2@yu.edu}
\affiliation{Department of Physics, Yeshiva University, New York, New York 10016, USA}
\author{Marcelo S. Sarandy}
\email{msarandy@if.uff.br}
\affiliation{Instituto de F\'{i}sica, Universidade Federal Fluminense, Av. Gal. Milton Tavares de Souza s/n, Gragoat\'{a}, 24210-346 Niter\'{o}i, Rio de Janeiro, Brazil}

\begin{abstract}
We investigate the phenomenon of spatial many-body localization (MBL) through pairwise correlation measures based on one and two-point correlation functions. The system considered is the Heisenberg spin-1/2 chain with exchange interaction $J$ and random onsite disorder of strength $h$.  As a representative pairwise correlation measure obtained from one-point functions only, we use global entanglement.  Through its finite size scaling analysis, we locate the MBL critical point at $h_{c}/J = 3.8$. As for measures involving two-point functions, we analyze pairwise geometric classical, quantum, and total correlations. Similarly to what happens for continuous quantum phase transitions, it is the derivatives of these two-point correlation measures that identify the MBL critical point, which is found to be in the range $h_{c}/J \in \left[3,4\right]$. Our approach relies on very simple measures that do not require access to multipartite entanglement or large portions of the system. 
\end{abstract}
\maketitle

\section{Introduction}

Concepts and tools from quantum information theory have found applications in the development of new numerical methods~\cite{Schollwock2011} and in fields as diverse as metrology~\cite{Toth2014}, high energy physics~\cite{Hayden2007}, and condensed matter physics~\cite{Amico2008,Li2008}. In particular, the successful detection of quantum phase transitions via quantum correlation measures, including concurrence~\cite{Osterloh2002,Wu:04}, entanglement entropy~\cite{Korepin:04}, and  quantum discord~\cite{Sarandy:09,Dillenschneider2008}, has motivated the use of these quantities in  theoretical~\cite{SantosEscobar2004,Brown2008,Dukesz2009,Mejia2005,Bera2016,Kjall2014,Yang2015,Goold2015,Tomasi2017}  and experimental~\cite{Smith2015,WeiARXIV} studies of the transition to many-body localized phases.

The term ``many-body localization'' (MBL) usually refers to spatial localization of interacting systems in the presence of onsite disorder. In one-dimensional noninteracting quantum systems, the inclusion of uncorrelated~\cite{Anderson1958} or quasiperiodic~\cite{Aubry1980}  onsite disorder takes the system into an insulating phase. In interacting systems, the interplay between interaction and disorder can cause the onset of quantum chaos~\cite{Avishai2002,Santos2004}, which greatly enhances delocalization. In these chaotic quantum systems with many interacting particles, the eigenstates away from the edges of the spectrum approach random vectors, therefore enabling the emergence of thermalization~\cite{Jensen1985,ZelevinskyRep1996,Borgonovi2016,Alessio2016}. It is natural to wonder whether such complex systems, capable of exhibiting statistical behavior despite isolation, may still become spatially localized under finite values of disorder strength. 

The viability of MBL was discussed already in Refs.~\cite{Anderson1958,Fleishman1980,Giamarchi1988} and perturbative approaches were employed to show that it indeed occurs at low temperatures~\cite{Gornyi2005,Basko2006}. For highly excited states, analytical~\cite{Aleiner2010,Imbrie2016a}, numerical~\cite{SantosEscobar2004,Brown2008,Dukesz2009,Kjall2014,Yang2015,Goold2015,Bera2016,Tomasi2017, Santos2005loc,Oganesyan2007,Znidaric2008,Pal2010,Canovi2011,Bardarson2012,DeLuca2013,Huse2014,Lev2014,GroverARXIV,Serbyn2014,Chandran2015,Luitz2015,Luitz2016,Torres2015,Torres2016BJP,Altman2015,Nandkishore2015,Serbyn2015,Singh2016,Agarwal2015,Bertrand2016,Gopalakrishnan2016,Monthus2016,Gogolin2016,Barisic2016,Khemani2016,Enss2017,Torres2017,TorresARXIV,LuitzARXIV}, and experimental~\cite{Smith2015,Schreiber2015,Kondov2015,WeiARXIV} studies also point toward a positive answer.  The numerical characterization of the transition to the MBL phase have included the analysis of level statistics~\cite{Santos2004,SantosEscobar2004,Brown2008,Dukesz2009,Oganesyan2007,Canovi2011,Bertrand2016} and its dynamical consequences~\cite{TorresARXIV}, delocalization measures~\cite{SantosEscobar2004,Brown2008,Dukesz2009,Pal2010,DeLuca2013,Torres2017}, transport properties~\cite{Barisic2016}, entanglement spectrum~\cite{Yang2015}, as well as power-law decays of the survival probability~\cite{Torres2015} and few-body observables~\cite{Serbyn2014,Luitz2016}.

From the point of view of quantum information theory, studies of MBL have taken several approaches. It has been shown, for instance, that deep in the chaotic phase, multipartite entanglement is large~\cite{Brown2008}, while a shift to pairwise entanglement takes place in the vicinity of the MBL transition~\cite{SantosEscobar2004,Brown2008,Dukesz2009,Mejia2005,Bera2016}. In the localized phase, the entanglement between two sites~\cite{Pal2010} and the quantum mutual information between two traced-out regions~\cite{Tomasi2017} fall off exponentially with their distance. As for the search for the critical point, scaling analysis of total multipartite correlations~\cite{Goold2015} and of the entanglement entropy~\cite{Nandkishore2015,Luitz2015} have been used. The latter has actually become the most popular method to characterize the MBL transition. In the chaotic (thermal) phase, for any bipartition of the chain, the scaling of the entanglement entropy for the  many-body eigenstates away from the edges of the spectrum obeys a volume law, while in the MBL phase, it exhibits an area law. In terms of dynamics, for initial states corresponding to computational basis vectors, as the system approaches the MBL phase, the entanglement entropy~\cite{Znidaric2008,Bardarson2012,Kjall2014} as well as  the Shannon information entropy~\cite{Torres2017} and the quantum mutual information~\cite{Tomasi2017}  grow logarithmically in time, which contrasts with the very fast increase and quick saturation in the chaotic regime. 

In the present work, instead of manipulating large parts of the quantum system, as in the case of bipartite blocks, or dealing with multipartite measures, we focus on the characterization of the MBL critical point through pairwise correlation measures that involve one or at most two-point correlation functions. Specifically, we consider global entanglement~\cite{Meyer:02,Brennen:03} and pairwise geometric correlations beyond entanglement, as defined in~\cite{Paula:13-1,Nakano:13,Paula:13-3}.
These simple pairwise correlation measures are accessible through quantum tomography, with one- and two-point functions readily provided by current experiments~\cite{Walborn:06,Jurcevic:14,Fukuhara:15}.

Global entanglement is based on one-point  functions only. It can be understood as an average pairwise correlation measure, where the pair consists of a single site and all the rest of the system. By collapsing the global entanglement curves for different system sizes onto a single scaling function, we are able to precisely identify the critical point.

Pairwise correlation measures beyond entanglement were first proposed in Refs.~\cite{Ollivier:01,Henderson2001}. In this scenario, to classify and quantify physical correlations, one separates the states into quantum and classical, rather than entangled and unentangled. Classical states are defined as those left undisturbed by a non-selective local measurement~\cite{Luo:08,Saguia:11}, while the opposite holds for quantum states. Not all separable states are classical, some exhibit quantum correlations~\cite{Ollivier:01,Henderson2001}. 
The latter are useful tools in the analysis of critical phenomena due to their robustness at finite temperature~\cite{Werlang:10} and their long-range behavior in critical phases~\cite{Maziero:10}. 

We investigate pairwise classical, quantum, and total correlations at the MBL transition. To evaluate them, we take a geometric approach, via the trace norm in state space~\cite{Paula:13-1,Nakano:13,Paula:13-3}. The use of geometric correlations is convenient, because they can be analytically derived for various classes of two-qubit states. We show that, similarly to what happens for ordinary continuous quantum phase transitions, it is the first derivative of these two-point correlation measures that detects the critical point.

\section{Heisenberg spin-1/2 chains in a random magnetic field}

We consider a closed isotropic Heisenberg spin-1/2 chain with $N$ sites and random static magnetic fields in the $z$-direction. The Hamiltonian is given by
\begin{equation}
H = \sum_{k=1}^{N}\left[J\left(S_{k}^{x}S_{k+1}^{x}+S_{k}^{y}S_{k+1}^{y}+S_{k}^{z}S_{k+1}^{z}\right)+h_{k}S_{k}^{z}\right] ,
\label{ham}
\end{equation}
where $\hbar=1$ and $S_{k}^{x,y,z} = \sigma_{k}^{x,y,z}/2$, with $\sigma^{x,y,z}_{k}$ denoting the Pauli operators on site $k$.  The Zeeman splittings $h_{k}$ are random numbers from a uniform distribution in the interval $\left[-h,h\right]$ and $J$ is the exchange interaction. Borrowing the language from quantum information theory, we refer to a spin-1/2 as a quantum bit (qubit). 

The model conserves total magnetization in the $z$-direction, ${\cal S}^z = \sum_k S_{k}^{z}$, that is $\left[H, {\cal S}^z \right]=0$. Our studies focus on the largest subspace, ${\cal S}^z=0$, where localization is more demanding. We exactly diagonalize the Hamiltonian matrix of this block. Our analysis is carried out for $10\%$ of the eigenstates that belong to the middle of the energy spectrum, where they tend to be more delocalized. We perform averages over these states and over disorder configurations. For $N = 10,12,14$, we average over $10^4$ disorder realizations, while for $N = 16$ we use $10^3$ configurations.

The Hamiltonian in Eq.~(\ref{ham}) has two integrable limits, one when the chain is clean, $h/J\!\!\!=\!\!\!0$, and the other for $h/J>h_c/J$, where $h_c/J$ is the MBL critical point. Previous works have estimated $h_c/J\sim 3.8$ \cite{Luitz2015,Nandkishore2015,Goold2015}. Between the two integrable regions, for $0<h/J<h_c/J$, the system shows level repulsion and intermediate level spacing distributions between Poisson (usual in integrable models) and Wigner-Dyson (typical of chaotic systems). For $N=16$, the best agreement with the Wigner-Dyson distribution, indicating that the system is deep in the chaotic regime, occurs for $h/J\sim0.5$ \cite{Torres2017}.

\section{Global entanglement}

Global entanglement, $G_{E}$, was originally introduced 
in Ref.~\cite{Meyer:02} as an entanglement measure for pure composite states that vanishes if, and only if, the state is a tensor product of all of its subsystems. As shown in Ref.~\cite{Brennen:03}, the measure can be expressed in terms of an average over pairwise entanglement between one site and the rest of the system. More specifically, $ G_{E}(|\psi\rangle)$ is obtained from the one-qubit reduced density operators as
\begin{equation}
 G_{E}(|\psi\rangle) = 2-\frac{2}{N}\sum_{k=1}^{N}\textrm{Tr}\left(\rho_{k}^{2}\right) .
 \label{eqgent}
\end{equation}
If translation invariance is obeyed,  Eq.~(\ref{eqgent}) reduces to
\begin{equation}
 G_{E}(|\psi\rangle) = 2\left[1- \textrm{Tr}\left(\rho_{1}^{2}\right)\right]\ , 
 \label{eqgent2}
\end{equation}
where $\rho_{1} = \rho_{k}$, $\forall k$. Equation~(\ref{eqgent2}) is the linear entropy for a single spin of the system. 

Global entanglement identifies the multipartite separability of pure states, but it is still a bipartite measure, as provided by Eq.~(\ref{eqgent}). It is in this sense that we say that the measures adopted in this work do not require access to multipartite entanglement. General properties of $G_E$ in random localized states disordered systems were considered in Ref.~\cite{Giraud:07}. The dependence of global entanglement on the disorder strength  was studied also in \cite{Dukesz2009}, although the characterization of the MBL critical point was not provided. 

\begin{figure}[htb]
\includegraphics[scale=0.34]{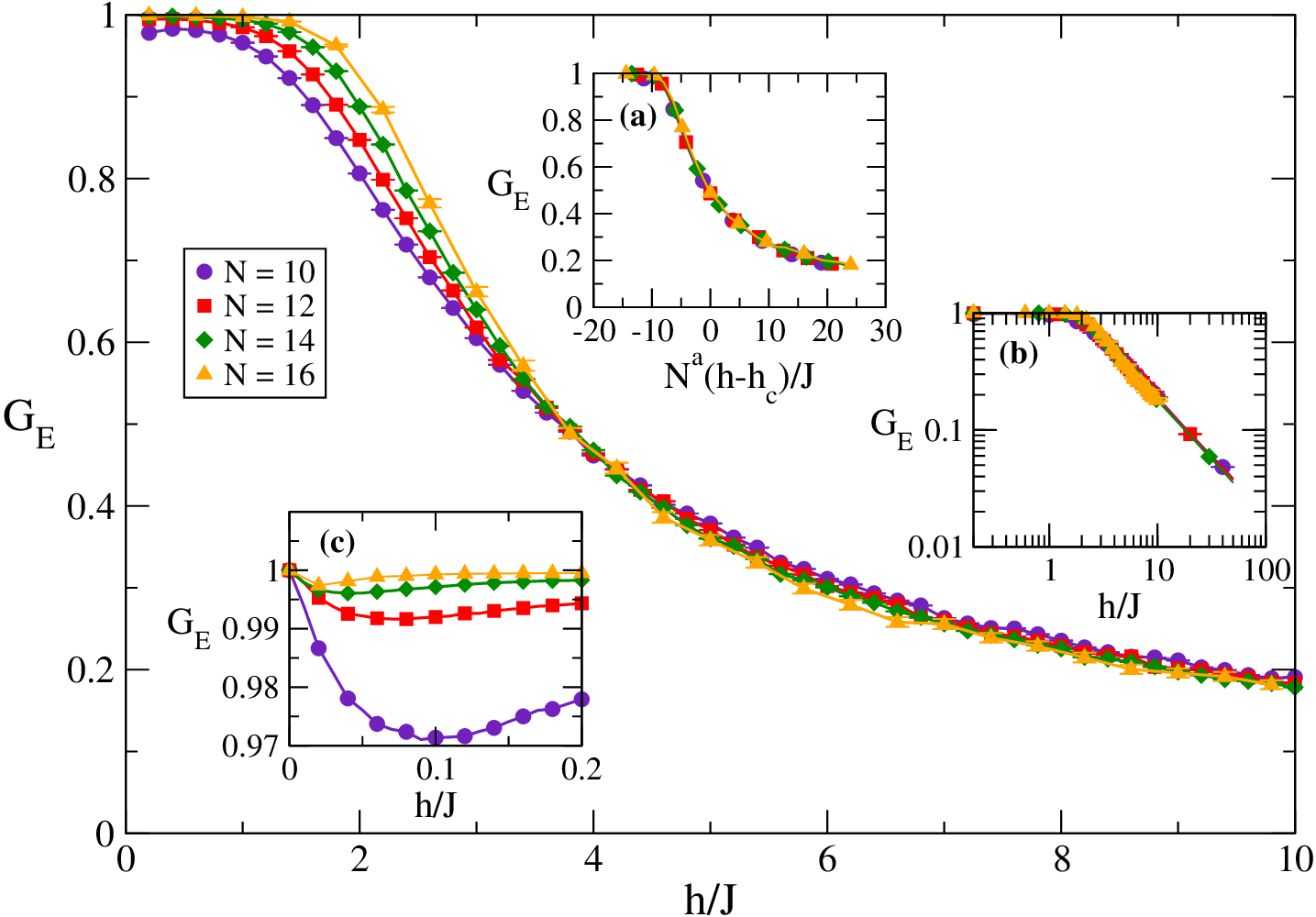}
\caption{(Color online) Main panel: Global entanglement $G_E$ as a function of the disorder strength $h/J$. 
Inset (a) shows the finite size scaling analysis, with $a = 0.5$ and $h_{c}/J = 3.8 \pm 0.2$. Inset (b) shows the 
power-law decay of $G_E$ in the localized phase. Inset (c) shows the non-monotonic behavior of $G_E$ for 
small values of $h/J$. }
\label{gent}
\end{figure}

The analysis of the MBL transition via global entanglement can be done directly with Eq.(\ref{eqgent2}), because the average over disorder realizations implies that all sites are equivalent.
Our results are shown in Fig.~\ref{gent}. As $h/J$ increases and the system approaches the MBL phase, the value of global entanglement decreases, since more entanglement gets localized in smaller subsystems, such as in pair of spins. 

The crossing of the curves in the main panel of Fig.~\ref{gent} indicates the approximate value of the critical point, which is then precisely obtained through a finite size scaling analysis. This is done by choosing the scaling form 
\begin{equation}
G_E =  \Phi\left[ N^a \left( \frac{h - h_c}{J} \right) \right],
\end{equation}
where $\Phi$ is a function determined by the chi-square minimization method. This function is employed in the scaling analysis presented in the inset (a) of Fig.~\ref{gent}. We find that $a=0.5$ and $h_{c}/J = 3.8 \pm 0.2$. This value of the critical point is in perfect agreement with previous studies~\cite{Luitz2015,Nandkishore2015,Goold2015}. It is remarkable that such a simple quantity, which corresponds to the entropy of a single spin, can determine so well the critical point. 

In the inset (b) of Fig.~\ref{gent}, we show the behavior of $G_E$ deep in the localized phase in a 
logarithmic scale. Note that it exhibits a power-law decay, $G_E \propto (h/J)^{-\alpha}$, with exponent 
roughly given by $\alpha=1$. This behavior closely resembles the multipartite mutual information, as discussed in Ref.~\cite{Goold2015}. 

We also notice that $G_E$ presents a non-monotonic behavior for small values of $h/J$. 
This is shown in the inset (c) of Fig.~\ref{gent}. The maximal value, 
$G_E=1$, occurs in the clean integrable limit, as originally shown in Ref.~\cite{Meyer:02}. As $h/J$ increases from zero, $G_E$ shows a minor dip, before reaching very large values again in the chaotic regime. Analogously to what happens to quantities that measure the integrable-chaos transition~\cite{Torres2017}, the dip shifts towards smaller $h/J$'s as the system size increases. This suggests that, in addition to identifying the MBL critical point, $G_E$ may be a general integrable-chaos detector.

\section{Geometric correlations for two-qubit states}

Pairwise correlations are analyzed in a bipartite Hilbert space ${\cal H} = {\cal H}_{1} \otimes {\cal H}_{2}$. The quantum states of the composite system are described by density operators $\rho \in {\cal B}({\cal H})$, where ${\cal B}({\cal H})$ is the set of bound, positive-semidefinite operators acting on ${\cal H}$ with trace given by ${\textrm{Tr}}\,\rho=1$. To distinguish classical from quantum correlation measures, we use the concept of classicality in quantum information. A state is classical if it is not disturbed under projective measurements~\cite{Luo:08}. 
Let us denote a set of local von Neumann measurements as $\{\Pi_{1}^{j} \otimes \mathbbm{1}_2\}$, 
where $\Pi_{1}^{j}$ is a set of orthogonal projectors over the first subsystem of the bipartition. 
After a non-selective measurement $M$, the density operator $\rho$ becomes
\begin{equation}
M(\rho) = \sum_j [\Pi_{1}^{j} \otimes \mathbbm{1}_2]\, \rho\, [\Pi_{1}^{j} \otimes \mathbbm{1}_2] .
\end{equation}
If there exists any measurement $M$ such that $M(\rho)=\rho$, then $\rho$ describes a hybrid state, referred to as classical-quantum state, which is classical with respect to the Hilbert space $ {\cal H}_{1} $ and potentially quantum with respect to  $ {\cal H}_{2}$. Extension of this definition for measurements over all subsystems and for multipartite states can be found in~\cite{Saguia:11}. 
We emphasize that separable mixed states are not necessarily classical and may still present quantum correlations. Therefore, non-classical states are not necessarily quantum entangled.

We define pairwise correlation measures by adopting a geometric approach based on the 
general formalism introduced in Refs.~\cite{Modi:12,Brodutch:12,Paula:14}.  We consider 
correlations based on the trace norm (Schatten 1-norm) and projective measurements 
operating over ${\cal H}_{1}$. The three kinds of correlations analyzed here are pairwise geometric classical, quantum, and total.

The amount of quantum correlation is measured through the geometric quantum discord, $Q_G(\rho)$,  defined as 
\begin{equation}
Q_G(\rho)=\min_{\{M\}}\,\mathrm{Tr}\left|\rho-M(\rho)\right|.
\end{equation}
The minimization is taken over all local measurements $M$ acting on ${\cal H}_{1}$. Thus, $Q_G(\rho)$ represents the distance between $\rho$ and 
the closest classical-quantum state obtained by measuring $\rho$.

The amount of classical correlation $C_G(\rho)$ associated with $\rho$ is obtained from
\begin{equation}
C_G(\rho)=\max_{\{\bar{M}\}}\,\mathrm{Tr}\left|\bar{M}(\rho)-\bar{M}(\pi_{\rho})\right|,
\label{cg-0}
\end{equation}
with the maximization done over all local measurements $\bar{M}$  acting on ${\cal H}_{1}$ and 
$\pi_{\rho}=\rho_1\otimes \rho_2 = \text{Tr}_{2}\rho\otimes\text{Tr}_{1}\rho$. 
To avoid ambiguities in the correlation measures for $Q_G$ and $C_G$, we take 
$M$ and $\bar{M}$ as independent measurement sets~\cite{Paula:14}. 
The classical 
correlation $C_G(\rho)$ relates to the maximum information about the state 
that we can locally extract by 
measuring $\rho$. It vanishes if and only if $\rho$ is a tensor product of its marginals, that is, if $\rho$ is the completely uncorrelated state $\pi_{\rho}$. 

For the total correlation, we simply take the 
trace distance between $\rho$ and $\pi_{\rho}$, which yields
\begin{equation}\label{tg}
T_G(\rho)= \mathrm{Tr}\left|\rho-\pi_{\rho}\right|.
\end{equation}  
The total correlation as provided by Eq.~(\ref{tg}) detects any kind of correlation that makes 
$\rho$ distinct from the trivial product state $\pi_\rho$. 

We study the correlations between two spins only and choose nearest-neighboring ones. The two-qubit states are described by the reduced density operator $\rho$ obtained after tracing out all spins of the chain except the two selected ones. For Hamiltonians that commute with the parity operator $\otimes_{i=1}^{N} S_i^z$, as $H$  in Eq.~(\ref{ham}), the reduced density matrix written in the computational basis $\left\{|00\rangle,|01\rangle,|10\rangle,|11\rangle\right\}$ has the form
\begin{equation}
\rho =\left(
\begin{array}{cccc}
\rho_{11} & 0 & 0 & \rho_{41}^{*} \\
0 & \rho_{22} & \rho_{32}^{*} & 0 \\
0 & \rho_{32} & \rho_{33} & 0 \\
\rho_{41} & 0 & 0 & \rho_{44} \\
\end{array}
\right) \ ,
\label{state}
\end{equation}
where the following constraints are assumed: $\sum_{i=1}^{4} \rho_{ii} = 1$ 
(normalization condition), $\rho_{11}\rho_{44} \geq |\rho_{41}|^{2}$, and 
$ \rho_{22}\rho_{33} \geq |\rho_{32}|^{2}$ (positive semidefiniteness). 
The nonzero elements $\rho_{ij}$ appear only in the diagonal and anti-diagonal of the reduced density matrix, which justifies the label ``X-state''. 
For this kind of two-qubit states, there are analytical expressions to calculate 
the geometric correlations~\cite{Ciccarello:14,Obando:15}. 
Indeed, defining the following 
auxiliary parameters 
$$c_{1}=2(\rho_{32}+\rho_{41}), \hspace{0.1 cm} c_{2}=2(\rho_{32}-\rho_{41}), $$ 
$$c_{3}=1-2(\rho_{22}+\rho_{33}), \hspace{0.1 cm} c_{4}=2(\rho_{11}+\rho_{33})-1,$$
$$c_{5}=2(\rho_{11}+\rho_{22})-1,$$
the geometric quantum discord is given by~\cite{Ciccarello:14}
\begin{equation}
 Q_{G}(\rho) = \sqrt{\frac{ac-bd}{a-b+c-d}}\ ,
\end{equation}
where $a = \max(c_{3}^{2},d+c_{5}^{2})$, $b = \min(c,c_{3}^{2})$, $c = \max(c_{1}^{2},c_{2}^{2})$ and 
$d = \min(c_{1}^{2},c_{2}^{2})$. The geometric classical and total correlations are respectively written as~\cite{Obando:15}
\begin{eqnarray}
 C_{G}(\rho) &=& c_{\max}, \\
 T_{G}(\rho) &=& \frac{1}{2}\left[c_{\max} + \max(c_{\max},c_{\textrm{int}}+c_{\min})\right]\ ,
\end{eqnarray}
where $c_{\max} = \max\left(|c_{1}|,|c_{2}|,|c_{3}-c_{4}c_{5}|\right)$, 
$c_{\min}= \min\left(|c_{1}|,|c_{2}|,|c_{3}-c_{4}c_{5}|\right)$, and $c_{\textrm{int}}= \textrm{int}\left(|c_{1}|,|c_{2}|,|c_{3}-c_{4}c_{5}|\right)$ corresponds to the intermediate value. 

\begin{figure}[!htb]
\includegraphics[scale=0.34]{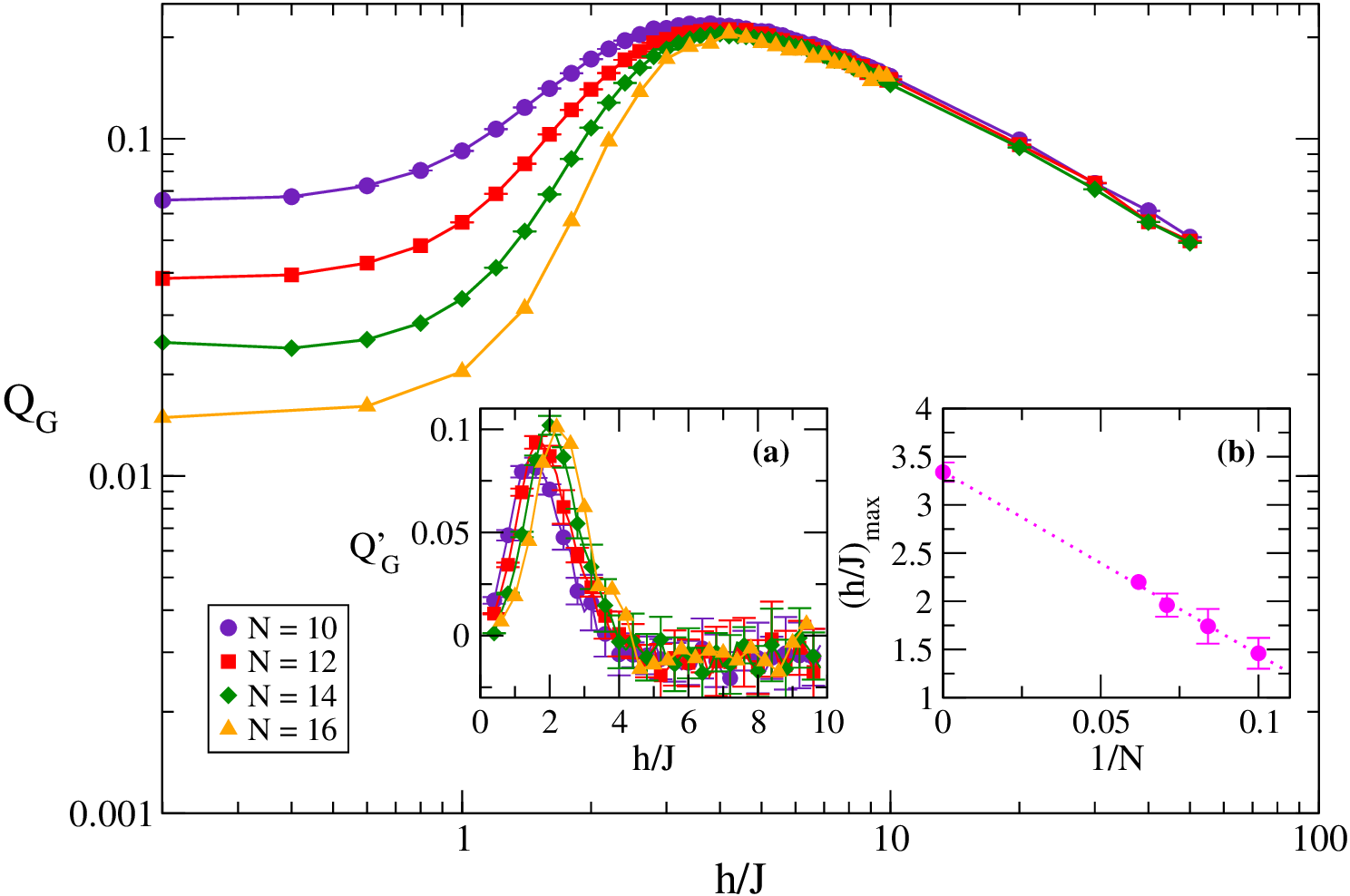}
\caption{(Color online) Main panel: Geometric quantum discord $Q_{G}$ as function of the disorder strength $h/J$ in 
a logarithmic scale. Inset (a): first derivative $Q^\prime_{G}$ of the geometric quantum discord 
with respect to $h/J$. Inset (b): the thermodynamic limit of the maximum of  $Q^\prime_{G}$, 
which yields $h_{c}/J = 3.34 \pm 0.03$.}
\label{qdis}
\end{figure}

\begin{figure}[!htb]
\includegraphics[scale=0.33]{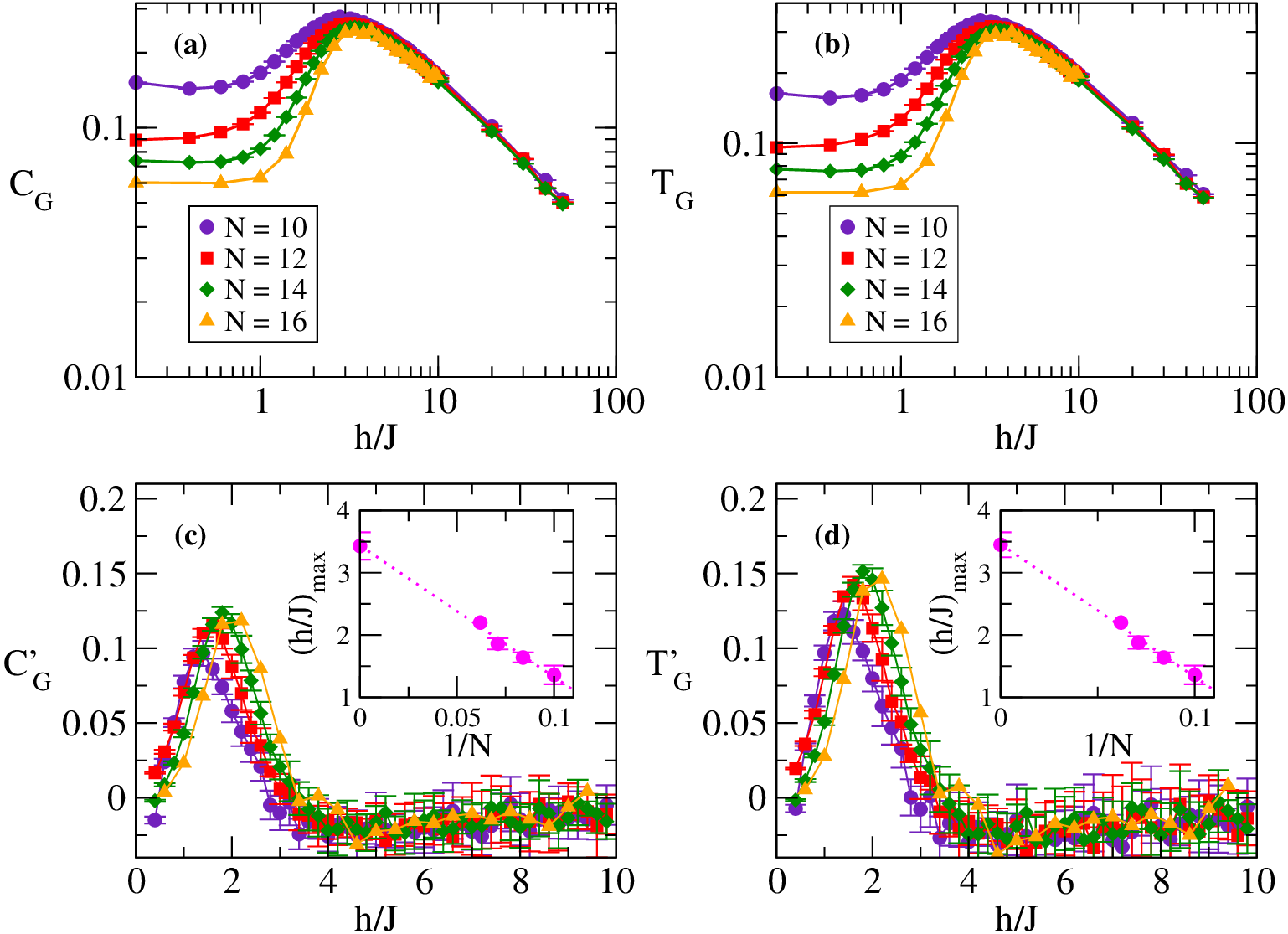}
\caption{(Color online) Geometric classical $C_{G}$ (a) and total $T_{G}$ (b) correlations as functions 
of $h/J$ in logarithmic scale, as well as their derivatives, $C_{G}^\prime$ (c)  and  $T_{G}^\prime$ (d), 
with respect to $h/J$. The insets in (c) and (d) exhibit the thermodynamic limit of the maxima of 
$C_{G}^\prime$ and $T_{G}^\prime$, respectively, yielding 
$h_{c}/J = 3.43 \pm 0.06$ for $C_{G}$ and $h_{c}/J = 3.45 \pm 0.06$ for $T_{G}$.}
\label{toge}
\end{figure}

The geometric quantum discord $Q_{G}$ between 
two nearest-neighboring spins is shown in the main panel of Fig.~\ref{qdis} as a function of $h/J$ for different chain sizes. The classical and total correlations, $C_G$ and $T_G$, are displayed in Figs.~\ref{toge} (a) and \ref{toge}  (b). The curves for the quantum, classical, and total correlations exhibit a similar pattern. They are non-monotonic and generalize the behavior of pairwise entanglement, as measured by 
concurrence~\cite{SantosEscobar2004,Bera2016}. We identify the following three regions described below.


In the chaotic phase, the correlations are spread out in the system in a multipartite form~\cite{Brown2008}. This results in a small concentration of correlations between individual pairs of spins and explains the small values of $Q_{G}$, $C_G$, and $T_G$. 

As the disorder strength increases beyond the chaotic region, $h/J>1$, one sees that $Q_{G}$, $C_G$, and $T_G$ increase up to a value of $h/J$ that depends on the length of the system. In the vicinity of the MBL transition, correlations are mostly confined between individual pairs of spins, which gives rise to the large values of  the quantum, classical, and total pairwise correlations.

After the transition to the MBL phase, further increasing the disorder strength asymptotically decreases the correlation measures due to the reduction of the effective role played by the exchange interaction $J$. As it happens for the global entanglement, the correlations $Q_{G}$, $C_G$ and $T_G$ show a power-law decay $\propto (h/J)^{-\beta}$ deep in the localized phase, with 
a universal exponent roughly given by $\beta = 0.7$. 
For a finite system, pairwise correlations tend to disappear in the limit of infinite disorder.  

The qualitative behavior of the quantum, classical, and total correlation measures discussed above already suggests the existence of a localized phase. 
The location of the critical point can be achieved by analyzing the first derivative of the correlation measures with respect to the 
disorder strength $h/J$. 
This approach is inspired by
the procedure used in studies of ordinary quantum phase transitions, either via pairwise entanglement~\cite{Osterloh2002,Wu:04} or via pairwise quantum discord~\cite{Sarandy:09,Dillenschneider2008}. 

As illustrated in the inset (a) of Fig.~\ref{qdis} and in Figs.~\ref{toge} (c) and~\ref{toge} (d), the three correlation measures exhibit a maximum in their derivatives, which occurs at a value of $h/J$ denoted by $(h/J)_{\text{max}}$. 
The maxima of the derivatives of the correlation measures obey a linear decay with $1/N$, from where the critical point can be obtained by extrapolating $(h/J)_{\text{max}}$ for 
$N\rightarrow \infty$.  

\begin{table}[!htb]
\centering
\begin{tabular}{|c|c|}
\hline
Correlation measure& $h_{c}/J$\\ \hline\hline
$G_E$ & $3.8 \pm 0.2$ \\
$Q_{G}$ & $3.34 \pm 0.03$\\
$C_{G}$ & $3.43 \pm 0.06$\\
$T_{G}$ & $3.45 \pm 0.06$\\
\hline
\end{tabular}
\caption{Critical point obtained via global entanglement in comparison with quantum, classical, and total geometric correlations. Error bars account only for fluctuations with respect to the number disorder realizations.  }
\label{table}
\end{table}
The scaling analysis of $(h/J)_{\text{max}}$, shown in the inset (b) of Fig.~\ref{qdis} and in the insets of Figs.~\ref{toge} (c) and~\ref{toge} (d), leads to $h_{c}/J \in \left[3,4\right]$, which coincides with the range of values for the critical point found in previous works. 
 Table~\ref{table} provides the values of the critical point extracted from each geometric correlation measure and also from the global entanglement $G_E$. Notice that the values are compatible with each other, which  indicates that the picture 
of delocalization of correlations in the thermal phase against concentration of correlations in the vicinity of the MBL 
critical point holds for quantum, classical, and 
total correlations between spin pairs.

\section{Conclusions}

We have described the transition to the MBL phase and located its critical point with pairwise correlation measures that involve one or at most two-point correlation functions. 
The results converged by taking only a small fraction of the eigenstates of the Hamiltonian.
Our main findings are:

(i) The finite size scaling analysis of the global entanglement led to the critical point $h_c/J = 3.8\pm 0.2$, which coincides with the result from previous works. This quantity, as computed here, is the linear entropy of a single spin. It is impressive that a simple one-point correlation function can be so effective at identifying the critical point. 

(ii) The geometric correlations between only two spins located the critical point in the range of $h_{c}/J \in \left[3,4\right]$, which is also within acceptable values. The procedure used resembles the one adopted in the analysis of ordinary quantum phase transitions, where the critical point is revealed by the derivative of the correlation measures. The scaling analysis of the derivatives of the quantum, classical, and total correlations followed the universal linear scaling law, that is, the derivatives are linear functions of $1/N$. The critical point is obtained by the extrapolation to the thermodynamic limit.

As a byproduct of our analysis, we found a non-monotonic behavior of the global entanglement in the transition from the clean integrable limit to the chaotic regime. A detailed characterization of the onset of chaos in terms of pairwise correlations is left as a plan for further studies. In the future, it might also be interesting to explore the effectiveness of the one and two-point correlation functions in the detection of the metal-insulator transition in higher dimensions. Another aspect that we intend to investigate is the changes that should be brought to the picture developed here when time-dependent Hamiltonians or the presence of decoherence are taken into account.

\section*{Acknowledgments}
J.L.C.C.F. is supported by CAPES-Brazil. 
A.S. thanks Stephan Haas for the use of the HPC cluster and for his hospitality at the University 
of Southern California, where this work initiated. L.F.S. is supported by the NSF grant No.~DMR-1603418. 
M.S.S. acknowledges support from CNPq-Brazil 
(No.303070/2016-1), FAPERJ (No 203036/2016), and the Brazilian National Institute 
for Science and Technology of Quantum Information (INCT-IQ).


\begin{thebibliography}{88}
\expandafter\ifx\csname natexlab\endcsname\relax\def\natexlab#1{#1}\fi
\expandafter\ifx\csname bibnamefont\endcsname\relax
  \def\bibnamefont#1{#1}\fi
\expandafter\ifx\csname bibfnamefont\endcsname\relax
  \def\bibfnamefont#1{#1}\fi
\expandafter\ifx\csname citenamefont\endcsname\relax
  \def\citenamefont#1{#1}\fi
\expandafter\ifx\csname url\endcsname\relax
  \def\url#1{\texttt{#1}}\fi
\expandafter\ifx\csname urlprefix\endcsname\relax\def\urlprefix{URL }\fi
\providecommand{\bibinfo}[2]{#2}
\providecommand{\eprint}[2][]{\url{#2}}

\bibitem[{\citenamefont{Schollw{\"o}ck}(2011)}]{Schollwock2011}
\bibinfo{author}{\bibfnamefont{U.}~\bibnamefont{Schollw{\"o}ck}},
  \bibinfo{journal}{Ann. of Phys.} \textbf{\bibinfo{volume}{326}},
  \bibinfo{pages}{96} (\bibinfo{year}{2011}).

\bibitem[{\citenamefont{T\'oth and Apellaniz}(2014)}]{Toth2014}
\bibinfo{author}{\bibfnamefont{G.}~\bibnamefont{T\'oth}} \bibnamefont{and}
  \bibinfo{author}{\bibfnamefont{I.}~\bibnamefont{Apellaniz}},
  \bibinfo{journal}{J. Phys. A} \textbf{\bibinfo{volume}{47}},
  \bibinfo{pages}{424006} (\bibinfo{year}{2014}).

\bibitem[{\citenamefont{Hayden and Preskill}(2007)}]{Hayden2007}
\bibinfo{author}{\bibfnamefont{P.}~\bibnamefont{Hayden}} \bibnamefont{and}
  \bibinfo{author}{\bibfnamefont{J.}~\bibnamefont{Preskill}},
  \bibinfo{journal}{JHEP} \textbf{\bibinfo{volume}{2007}}, \bibinfo{pages}{120}
  (\bibinfo{year}{2007}).

\bibitem[{\citenamefont{Amico et~al.}(2008)\citenamefont{Amico, Fazio,
  Osterloh, and Vedral}}]{Amico2008}
\bibinfo{author}{\bibfnamefont{L.}~\bibnamefont{Amico}},
  \bibinfo{author}{\bibfnamefont{R.}~\bibnamefont{Fazio}},
  \bibinfo{author}{\bibfnamefont{A.}~\bibnamefont{Osterloh}}, \bibnamefont{and}
  \bibinfo{author}{\bibfnamefont{V.}~\bibnamefont{Vedral}},
  \bibinfo{journal}{Rev. Mod. Phys.} \textbf{\bibinfo{volume}{80}},
  \bibinfo{pages}{517} (\bibinfo{year}{2008}).

\bibitem[{\citenamefont{Li and Haldane}(2008)}]{Li2008}
\bibinfo{author}{\bibfnamefont{H.}~\bibnamefont{Li}} \bibnamefont{and}
  \bibinfo{author}{\bibfnamefont{F.~D.~M.} \bibnamefont{Haldane}},
  \bibinfo{journal}{Phys. Rev. Lett.} \textbf{\bibinfo{volume}{101}},
  \bibinfo{pages}{010504} (\bibinfo{year}{2008}).

\bibitem[{\citenamefont{Osterloh et~al.}(2002)\citenamefont{Osterloh, Amico,
  Falci, and Fazio}}]{Osterloh2002}
\bibinfo{author}{\bibfnamefont{A.}~\bibnamefont{Osterloh}},
  \bibinfo{author}{\bibfnamefont{L.}~\bibnamefont{Amico}},
  \bibinfo{author}{\bibfnamefont{G.}~\bibnamefont{Falci}}, \bibnamefont{and}
  \bibinfo{author}{\bibfnamefont{R.}~\bibnamefont{Fazio}},
  \bibinfo{journal}{Nature} \textbf{\bibinfo{volume}{416}},
  \bibinfo{pages}{608} (\bibinfo{year}{2002}).

\bibitem[{\citenamefont{Wu et~al.}(2004)\citenamefont{Wu, Sarandy, and
  Lidar}}]{Wu:04}
\bibinfo{author}{\bibfnamefont{L.-A.} \bibnamefont{Wu}},
  \bibinfo{author}{\bibfnamefont{M.~S.} \bibnamefont{Sarandy}},
  \bibnamefont{and} \bibinfo{author}{\bibfnamefont{D.~A.} \bibnamefont{Lidar}},
  \bibinfo{journal}{Phys. Rev. Lett.} \textbf{\bibinfo{volume}{93}},
  \bibinfo{pages}{250404} (\bibinfo{year}{2004}).

\bibitem[{\citenamefont{Korepin}(2004)}]{Korepin:04}
\bibinfo{author}{\bibfnamefont{V.~E.} \bibnamefont{Korepin}},
  \bibinfo{journal}{Phys. Rev. Lett.} \textbf{\bibinfo{volume}{92}},
  \bibinfo{pages}{096402} (\bibinfo{year}{2004}).

\bibitem[{\citenamefont{Sarandy}(2009)}]{Sarandy:09}
\bibinfo{author}{\bibfnamefont{M.~S.} \bibnamefont{Sarandy}},
  \bibinfo{journal}{Phys. Rev. A} \textbf{\bibinfo{volume}{80}},
  \bibinfo{pages}{022108} (\bibinfo{year}{2009}).

\bibitem[{\citenamefont{Dillenschneider}(2008)}]{Dillenschneider2008}
\bibinfo{author}{\bibfnamefont{R.}~\bibnamefont{Dillenschneider}},
  \bibinfo{journal}{Phys. Rev. B} \textbf{\bibinfo{volume}{78}},
  \bibinfo{pages}{224413} (\bibinfo{year}{2008}).

\bibitem[{\citenamefont{Santos et~al.}(2004)\citenamefont{Santos, Rigolin, and
  Escobar}}]{SantosEscobar2004}
\bibinfo{author}{\bibfnamefont{L.~F.} \bibnamefont{Santos}},
  \bibinfo{author}{\bibfnamefont{G.}~\bibnamefont{Rigolin}}, \bibnamefont{and}
  \bibinfo{author}{\bibfnamefont{C.~O.} \bibnamefont{Escobar}},
  \bibinfo{journal}{Phys. Rev. A} \textbf{\bibinfo{volume}{69}},
  \bibinfo{pages}{042304} (\bibinfo{year}{2004}).

\bibitem[{\citenamefont{Brown et~al.}(2008)\citenamefont{Brown, Santos,
  Starling, and Viola}}]{Brown2008}
\bibinfo{author}{\bibfnamefont{W.~G.} \bibnamefont{Brown}},
  \bibinfo{author}{\bibfnamefont{L.~F.} \bibnamefont{Santos}},
  \bibinfo{author}{\bibfnamefont{D.}~\bibnamefont{Starling}}, \bibnamefont{and}
  \bibinfo{author}{\bibfnamefont{L.}~\bibnamefont{Viola}},
  \bibinfo{journal}{Phys. Rev. E} \textbf{\bibinfo{volume}{77}},
  \bibinfo{pages}{021106} (\bibinfo{year}{2008}).

\bibitem[{\citenamefont{Dukesz et~al.}(2009)\citenamefont{Dukesz, Zilbergerts,
  and Santos}}]{Dukesz2009}
\bibinfo{author}{\bibfnamefont{F.}~\bibnamefont{Dukesz}},
  \bibinfo{author}{\bibfnamefont{M.}~\bibnamefont{Zilbergerts}},
  \bibnamefont{and} \bibinfo{author}{\bibfnamefont{L.~F.}
  \bibnamefont{Santos}}, \bibinfo{journal}{New J. Phys.}
  \textbf{\bibinfo{volume}{11}}, \bibinfo{pages}{043026}
  (\bibinfo{year}{2009}).

\bibitem[{\citenamefont{Mejia-Monasterio
  et~al.}(2005)\citenamefont{Mejia-Monasterio, Benenti, Carlo, and
  Casati}}]{Mejia2005}
\bibinfo{author}{\bibfnamefont{C.}~\bibnamefont{Mejia-Monasterio}},
  \bibinfo{author}{\bibfnamefont{G.}~\bibnamefont{Benenti}},
  \bibinfo{author}{\bibfnamefont{G.}~\bibnamefont{Carlo}}, \bibnamefont{and}
  \bibinfo{author}{\bibfnamefont{G.}~\bibnamefont{Casati}},
  \bibinfo{journal}{Phys. Rev. A} \textbf{\bibinfo{volume}{71}},
  \bibinfo{pages}{062324} (\bibinfo{year}{2005}).

\bibitem[{\citenamefont{Bera and Lakshminarayan}(2016)}]{Bera2016}
\bibinfo{author}{\bibfnamefont{S.}~\bibnamefont{Bera}} \bibnamefont{and}
  \bibinfo{author}{\bibfnamefont{A.}~\bibnamefont{Lakshminarayan}},
  \bibinfo{journal}{Phys. Rev. B} \textbf{\bibinfo{volume}{93}},
  \bibinfo{pages}{134204} (\bibinfo{year}{2016}).

\bibitem[{\citenamefont{Kj\"all et~al.}(2014)\citenamefont{Kj\"all, Bardarson,
  and Pollmann}}]{Kjall2014}
\bibinfo{author}{\bibfnamefont{J.~A.} \bibnamefont{Kj\"all}},
  \bibinfo{author}{\bibfnamefont{J.~H.} \bibnamefont{Bardarson}},
  \bibnamefont{and} \bibinfo{author}{\bibfnamefont{F.}~\bibnamefont{Pollmann}},
  \bibinfo{journal}{Phys. Rev. Lett.} \textbf{\bibinfo{volume}{113}},
  \bibinfo{pages}{107204} (\bibinfo{year}{2014}).

\bibitem[{\citenamefont{Yang et~al.}(2015)\citenamefont{Yang, Chamon, Hamma,
  and Mucciolo}}]{Yang2015}
\bibinfo{author}{\bibfnamefont{Z.-C.} \bibnamefont{Yang}},
  \bibinfo{author}{\bibfnamefont{C.}~\bibnamefont{Chamon}},
  \bibinfo{author}{\bibfnamefont{A.}~\bibnamefont{Hamma}}, \bibnamefont{and}
  \bibinfo{author}{\bibfnamefont{E.~R.} \bibnamefont{Mucciolo}},
  \bibinfo{journal}{Phys. Rev. Lett.} \textbf{\bibinfo{volume}{115}},
  \bibinfo{pages}{267206} (\bibinfo{year}{2015}).

\bibitem[{\citenamefont{Goold et~al.}(2015)\citenamefont{Goold, Gogolin, Clark,
  Eisert, Scardicchio, and Silva}}]{Goold2015}
\bibinfo{author}{\bibfnamefont{J.}~\bibnamefont{Goold}},
  \bibinfo{author}{\bibfnamefont{C.}~\bibnamefont{Gogolin}},
  \bibinfo{author}{\bibfnamefont{S.~R.} \bibnamefont{Clark}},
  \bibinfo{author}{\bibfnamefont{J.}~\bibnamefont{Eisert}},
  \bibinfo{author}{\bibfnamefont{A.}~\bibnamefont{Scardicchio}},
  \bibnamefont{and} \bibinfo{author}{\bibfnamefont{A.}~\bibnamefont{Silva}},
  \bibinfo{journal}{Phys. Rev. B} \textbf{\bibinfo{volume}{92}},
  \bibinfo{pages}{180202} (\bibinfo{year}{2015}).

\bibitem[{\citenamefont{De~Tomasi et~al.}(2017)\citenamefont{De~Tomasi, Bera,
  Bardarson, and Pollmann}}]{Tomasi2017}
\bibinfo{author}{\bibfnamefont{G.}~\bibnamefont{De~Tomasi}},
  \bibinfo{author}{\bibfnamefont{S.}~\bibnamefont{Bera}},
  \bibinfo{author}{\bibfnamefont{J.~H.} \bibnamefont{Bardarson}},
  \bibnamefont{and} \bibinfo{author}{\bibfnamefont{F.}~\bibnamefont{Pollmann}},
  \bibinfo{journal}{Phys. Rev. Lett.} \textbf{\bibinfo{volume}{118}},
  \bibinfo{pages}{016804} (\bibinfo{year}{2017}).

\bibitem[{\citenamefont{Smith et~al.}(2016)\citenamefont{Smith, Lee, Richerme,
  Neyenhuis, Hess, Hauke, Heyl, Huse, and Monroe}}]{Smith2015}
\bibinfo{author}{\bibfnamefont{J.}~\bibnamefont{Smith}},
  \bibinfo{author}{\bibfnamefont{A.}~\bibnamefont{Lee}},
  \bibinfo{author}{\bibfnamefont{P.}~\bibnamefont{Richerme}},
  \bibinfo{author}{\bibfnamefont{B.}~\bibnamefont{Neyenhuis}},
  \bibinfo{author}{\bibfnamefont{P.~W.} \bibnamefont{Hess}},
  \bibinfo{author}{\bibfnamefont{P.}~\bibnamefont{Hauke}},
  \bibinfo{author}{\bibfnamefont{M.}~\bibnamefont{Heyl}},
  \bibinfo{author}{\bibfnamefont{D.~A.} \bibnamefont{Huse}}, \bibnamefont{and}
  \bibinfo{author}{\bibfnamefont{C.}~\bibnamefont{Monroe}},
  \bibinfo{journal}{Nat. Phys.} \textbf{\bibinfo{volume}{12}},
  \bibinfo{pages}{907} (\bibinfo{year}{2016}).

\bibitem[{\citenamefont{Wei et~al.}()\citenamefont{Wei, Ramanathan, and
  Cappellaro}}]{WeiARXIV}
\bibinfo{author}{\bibfnamefont{K.~X.} \bibnamefont{Wei}},
  \bibinfo{author}{\bibfnamefont{C.}~\bibnamefont{Ramanathan}},
  \bibnamefont{and}
  \bibinfo{author}{\bibfnamefont{P.}~\bibnamefont{Cappellaro}},
  \bibinfo{note}{arXiv:1612.05249}.

\bibitem[{\citenamefont{Anderson}(1958)}]{Anderson1958}
\bibinfo{author}{\bibfnamefont{P.~W.} \bibnamefont{Anderson}},
  \bibinfo{journal}{Phys. Rev.} \textbf{\bibinfo{volume}{109}},
  \bibinfo{pages}{1492} (\bibinfo{year}{1958}).

\bibitem[{\citenamefont{Aubry and Andr\'e}(1980)}]{Aubry1980}
\bibinfo{author}{\bibfnamefont{S.}~\bibnamefont{Aubry}} \bibnamefont{and}
  \bibinfo{author}{\bibfnamefont{G.}~\bibnamefont{Andr\'e}},
  \bibinfo{journal}{Ann. Isr. Phys. Soc.} \textbf{\bibinfo{volume}{3}},
  \bibinfo{pages}{1335} (\bibinfo{year}{1980}).

\bibitem[{\citenamefont{Avishai et~al.}(2002)\citenamefont{Avishai, Richert,
  and Berkovitz}}]{Avishai2002}
\bibinfo{author}{\bibfnamefont{Y.}~\bibnamefont{Avishai}},
  \bibinfo{author}{\bibfnamefont{J.}~\bibnamefont{Richert}}, \bibnamefont{and}
  \bibinfo{author}{\bibfnamefont{R.}~\bibnamefont{Berkovitz}},
  \bibinfo{journal}{Phys. Rev. B} \textbf{\bibinfo{volume}{66}},
  \bibinfo{pages}{052416} (\bibinfo{year}{2002}).

\bibitem[{\citenamefont{Santos}(2004)}]{Santos2004}
\bibinfo{author}{\bibfnamefont{L.~F.} \bibnamefont{Santos}},
  \bibinfo{journal}{J. Phys. A} \textbf{\bibinfo{volume}{37}},
  \bibinfo{pages}{4723} (\bibinfo{year}{2004}).

\bibitem[{\citenamefont{Jensen and Shankar}(1985)}]{Jensen1985}
\bibinfo{author}{\bibfnamefont{R.~V.} \bibnamefont{Jensen}} \bibnamefont{and}
  \bibinfo{author}{\bibfnamefont{R.}~\bibnamefont{Shankar}},
  \bibinfo{journal}{Phys. Rev. Lett.} \textbf{\bibinfo{volume}{54}},
  \bibinfo{pages}{1879} (\bibinfo{year}{1985}).

\bibitem[{\citenamefont{Zelevinsky et~al.}(1996)\citenamefont{Zelevinsky,
  Brown, Frazier, and Horoi}}]{ZelevinskyRep1996}
\bibinfo{author}{\bibfnamefont{V.}~\bibnamefont{Zelevinsky}},
  \bibinfo{author}{\bibfnamefont{B.~A.} \bibnamefont{Brown}},
  \bibinfo{author}{\bibfnamefont{N.}~\bibnamefont{Frazier}}, \bibnamefont{and}
  \bibinfo{author}{\bibfnamefont{M.}~\bibnamefont{Horoi}},
  \bibinfo{journal}{Phys. Rep.} \textbf{\bibinfo{volume}{276}},
  \bibinfo{pages}{85} (\bibinfo{year}{1996}).

\bibitem[{\citenamefont{Borgonovi et~al.}(2016)\citenamefont{Borgonovi,
  Izrailev, Santos, and Zelevinsky}}]{Borgonovi2016}
\bibinfo{author}{\bibfnamefont{F.}~\bibnamefont{Borgonovi}},
  \bibinfo{author}{\bibfnamefont{F.~M.} \bibnamefont{Izrailev}},
  \bibinfo{author}{\bibfnamefont{L.~F.} \bibnamefont{Santos}},
  \bibnamefont{and} \bibinfo{author}{\bibfnamefont{V.~G.}
  \bibnamefont{Zelevinsky}}, \bibinfo{journal}{Phys. Rep.}
  \textbf{\bibinfo{volume}{626}}, \bibinfo{pages}{1} (\bibinfo{year}{2016}).

\bibitem[{\citenamefont{D'Alessio et~al.}(2016)\citenamefont{D'Alessio, Kafri,
  Polkovnikov, and Rigol}}]{Alessio2016}
\bibinfo{author}{\bibfnamefont{L.}~\bibnamefont{D'Alessio}},
  \bibinfo{author}{\bibfnamefont{Y.}~\bibnamefont{Kafri}},
  \bibinfo{author}{\bibfnamefont{A.}~\bibnamefont{Polkovnikov}},
  \bibnamefont{and} \bibinfo{author}{\bibfnamefont{M.}~\bibnamefont{Rigol}},
  \bibinfo{journal}{Adv. Phys.} \textbf{\bibinfo{volume}{65}},
  \bibinfo{pages}{239} (\bibinfo{year}{2016}).

\bibitem[{\citenamefont{Fleishman and Anderson}(1980)}]{Fleishman1980}
\bibinfo{author}{\bibfnamefont{L.}~\bibnamefont{Fleishman}} \bibnamefont{and}
  \bibinfo{author}{\bibfnamefont{P.~W.} \bibnamefont{Anderson}},
  \bibinfo{journal}{Phys. Rev. B} \textbf{\bibinfo{volume}{21}},
  \bibinfo{pages}{2366} (\bibinfo{year}{1980}).

\bibitem[{\citenamefont{Giamarchi and Schulz}(1988)}]{Giamarchi1988}
\bibinfo{author}{\bibfnamefont{T.}~\bibnamefont{Giamarchi}} \bibnamefont{and}
  \bibinfo{author}{\bibfnamefont{H.~J.} \bibnamefont{Schulz}},
  \bibinfo{journal}{Phys. Rev. B} \textbf{\bibinfo{volume}{37}},
  \bibinfo{pages}{325} (\bibinfo{year}{1988}).

\bibitem[{\citenamefont{Gornyi et~al.}(2005)\citenamefont{Gornyi, Mirlin, and
  Polyakov}}]{Gornyi2005}
\bibinfo{author}{\bibfnamefont{I.~V.} \bibnamefont{Gornyi}},
  \bibinfo{author}{\bibfnamefont{A.~D.} \bibnamefont{Mirlin}},
  \bibnamefont{and} \bibinfo{author}{\bibfnamefont{D.~G.}
  \bibnamefont{Polyakov}}, \bibinfo{journal}{Phys. Rev. Lett.}
  \textbf{\bibinfo{volume}{95}}, \bibinfo{pages}{206603}
  (\bibinfo{year}{2005}).

\bibitem[{\citenamefont{Basko et~al.}(2006)\citenamefont{Basko, Aleiner, and
  Altshuler}}]{Basko2006}
\bibinfo{author}{\bibfnamefont{D.~M.} \bibnamefont{Basko}},
  \bibinfo{author}{\bibfnamefont{I.~L.} \bibnamefont{Aleiner}},
  \bibnamefont{and} \bibinfo{author}{\bibfnamefont{B.~L.}
  \bibnamefont{Altshuler}}, \bibinfo{journal}{Ann. Phys.}
  \textbf{\bibinfo{volume}{321}}, \bibinfo{pages}{1126} (\bibinfo{year}{2006}).

\bibitem[{\citenamefont{Aleiner et~al.}(2010)\citenamefont{Aleiner, Altshuler,
  and Shlyapnikov}}]{Aleiner2010}
\bibinfo{author}{\bibfnamefont{I.~L.} \bibnamefont{Aleiner}},
  \bibinfo{author}{\bibfnamefont{B.~L.} \bibnamefont{Altshuler}},
  \bibnamefont{and} \bibinfo{author}{\bibfnamefont{G.~V.}
  \bibnamefont{Shlyapnikov}}, \bibinfo{journal}{Nat. Phys.}
  \textbf{\bibinfo{volume}{6}}, \bibinfo{pages}{900} (\bibinfo{year}{2010}).

\bibitem[{\citenamefont{Imbrie}(2016)}]{Imbrie2016a}
\bibinfo{author}{\bibfnamefont{J.~Z.} \bibnamefont{Imbrie}},
  \bibinfo{journal}{J. Stat. Phys.} \textbf{\bibinfo{volume}{163}},
  \bibinfo{pages}{998} (\bibinfo{year}{2016}).

\bibitem[{\citenamefont{Santos et~al.}(2005)\citenamefont{Santos, Dykman,
  Shapiro, and Izrailev}}]{Santos2005loc}
\bibinfo{author}{\bibfnamefont{L.~F.} \bibnamefont{Santos}},
  \bibinfo{author}{\bibfnamefont{M.~I.} \bibnamefont{Dykman}},
  \bibinfo{author}{\bibfnamefont{M.}~\bibnamefont{Shapiro}}, \bibnamefont{and}
  \bibinfo{author}{\bibfnamefont{F.~M.} \bibnamefont{Izrailev}},
  \bibinfo{journal}{Phys. Rev. A} \textbf{\bibinfo{volume}{71}},
  \bibinfo{pages}{012317} (\bibinfo{year}{2005}).

\bibitem[{\citenamefont{Oganesyan and Huse}(2007)}]{Oganesyan2007}
\bibinfo{author}{\bibfnamefont{V.}~\bibnamefont{Oganesyan}} \bibnamefont{and}
  \bibinfo{author}{\bibfnamefont{D.~A.} \bibnamefont{Huse}},
  \bibinfo{journal}{Phys. Rev. B} \textbf{\bibinfo{volume}{75}},
  \bibinfo{pages}{155111} (\bibinfo{year}{2007}).

\bibitem[{\citenamefont{\ifmmode \check{Z}\else
  \v{Z}\fi{}nidari\ifmmode~\check{c}\else \v{c}\fi{}
  et~al.}(2008)\citenamefont{\ifmmode \check{Z}\else
  \v{Z}\fi{}nidari\ifmmode~\check{c}\else \v{c}\fi{}, Prosen, and
  Prelov\ifmmode~\check{s}\else \v{s}\fi{}ek}}]{Znidaric2008}
\bibinfo{author}{\bibfnamefont{M.}~\bibnamefont{\ifmmode \check{Z}\else
  \v{Z}\fi{}nidari\ifmmode~\check{c}\else \v{c}\fi{}}},
  \bibinfo{author}{\bibfnamefont{T.}~\bibnamefont{Prosen}}, \bibnamefont{and}
  \bibinfo{author}{\bibfnamefont{P.}~\bibnamefont{Prelov\ifmmode~\check{s}\else
  \v{s}\fi{}ek}}, \bibinfo{journal}{Phys. Rev. B}
  \textbf{\bibinfo{volume}{77}}, \bibinfo{pages}{064426}
  (\bibinfo{year}{2008}).

\bibitem[{\citenamefont{Pal and Huse}(2010)}]{Pal2010}
\bibinfo{author}{\bibfnamefont{A.}~\bibnamefont{Pal}} \bibnamefont{and}
  \bibinfo{author}{\bibfnamefont{D.~A.} \bibnamefont{Huse}},
  \bibinfo{journal}{Phys. Rev. B} \textbf{\bibinfo{volume}{82}},
  \bibinfo{pages}{174411} (\bibinfo{year}{2010}).

\bibitem[{\citenamefont{Canovi et~al.}(2011)\citenamefont{Canovi, Rossini,
  Fazio, Santoro, and Silva}}]{Canovi2011}
\bibinfo{author}{\bibfnamefont{E.}~\bibnamefont{Canovi}},
  \bibinfo{author}{\bibfnamefont{D.}~\bibnamefont{Rossini}},
  \bibinfo{author}{\bibfnamefont{R.}~\bibnamefont{Fazio}},
  \bibinfo{author}{\bibfnamefont{G.~E.} \bibnamefont{Santoro}},
  \bibnamefont{and} \bibinfo{author}{\bibfnamefont{A.}~\bibnamefont{Silva}},
  \bibinfo{journal}{Phys. Rev. B} \textbf{\bibinfo{volume}{83}},
  \bibinfo{pages}{094431} (\bibinfo{year}{2011}).

\bibitem[{\citenamefont{Bardarson et~al.}(2012)\citenamefont{Bardarson,
  Pollmann, and Moore}}]{Bardarson2012}
\bibinfo{author}{\bibfnamefont{J.~H.} \bibnamefont{Bardarson}},
  \bibinfo{author}{\bibfnamefont{F.}~\bibnamefont{Pollmann}}, \bibnamefont{and}
  \bibinfo{author}{\bibfnamefont{J.~E.} \bibnamefont{Moore}},
  \bibinfo{journal}{Phys. Rev. Lett.} \textbf{\bibinfo{volume}{109}},
  \bibinfo{pages}{017202} (\bibinfo{year}{2012}).

\bibitem[{\citenamefont{Luca and Scardicchio}(2013)}]{DeLuca2013}
\bibinfo{author}{\bibfnamefont{A.~D.} \bibnamefont{Luca}} \bibnamefont{and}
  \bibinfo{author}{\bibfnamefont{A.}~\bibnamefont{Scardicchio}},
  \bibinfo{journal}{Europhys. Lett.} \textbf{\bibinfo{volume}{101}},
  \bibinfo{pages}{37003} (\bibinfo{year}{2013}).

\bibitem[{\citenamefont{Huse et~al.}(2014)\citenamefont{Huse, Nandkishore, and
  Oganesyan}}]{Huse2014}
\bibinfo{author}{\bibfnamefont{D.~A.} \bibnamefont{Huse}},
  \bibinfo{author}{\bibfnamefont{R.}~\bibnamefont{Nandkishore}},
  \bibnamefont{and}
  \bibinfo{author}{\bibfnamefont{V.}~\bibnamefont{Oganesyan}},
  \bibinfo{journal}{Phys. Rev. B} \textbf{\bibinfo{volume}{90}},
  \bibinfo{pages}{174202} (\bibinfo{year}{2014}).

\bibitem[{\citenamefont{Bar~Lev and Reichman}(2014)}]{Lev2014}
\bibinfo{author}{\bibfnamefont{Y.}~\bibnamefont{Bar~Lev}} \bibnamefont{and}
  \bibinfo{author}{\bibfnamefont{D.~R.} \bibnamefont{Reichman}},
  \bibinfo{journal}{Phys. Rev. B} \textbf{\bibinfo{volume}{89}},
  \bibinfo{pages}{220201} (\bibinfo{year}{2014}).

\bibitem[{\citenamefont{Grover}()}]{GroverARXIV}
\bibinfo{author}{\bibfnamefont{T.}~\bibnamefont{Grover}},
  \bibinfo{note}{arXiv:1405.1471}.

\bibitem[{\citenamefont{Serbyn et~al.}(2014)\citenamefont{Serbyn,
  Papi\ifmmode~\acute{c}\else \'{c}\fi{}, and Abanin}}]{Serbyn2014}
\bibinfo{author}{\bibfnamefont{M.}~\bibnamefont{Serbyn}},
  \bibinfo{author}{\bibfnamefont{Z.}~\bibnamefont{Papi\ifmmode~\acute{c}\else
  \'{c}\fi{}}}, \bibnamefont{and} \bibinfo{author}{\bibfnamefont{D.~A.}
  \bibnamefont{Abanin}}, \bibinfo{journal}{Phys. Rev. B}
  \textbf{\bibinfo{volume}{90}}, \bibinfo{pages}{174302}
  (\bibinfo{year}{2014}).

\bibitem[{\citenamefont{Chandran and Laumann}(2015)}]{Chandran2015}
\bibinfo{author}{\bibfnamefont{A.}~\bibnamefont{Chandran}} \bibnamefont{and}
  \bibinfo{author}{\bibfnamefont{C.~R.} \bibnamefont{Laumann}},
  \bibinfo{journal}{Phys. Rev. B} \textbf{\bibinfo{volume}{92}},
  \bibinfo{pages}{024301} (\bibinfo{year}{2015}).

\bibitem[{\citenamefont{Luitz et~al.}(2015)\citenamefont{Luitz, Laflorencie,
  and Alet}}]{Luitz2015}
\bibinfo{author}{\bibfnamefont{D.~J.} \bibnamefont{Luitz}},
  \bibinfo{author}{\bibfnamefont{N.}~\bibnamefont{Laflorencie}},
  \bibnamefont{and} \bibinfo{author}{\bibfnamefont{F.}~\bibnamefont{Alet}},
  \bibinfo{journal}{Phys. Rev. B} \textbf{\bibinfo{volume}{91}},
  \bibinfo{pages}{081103} (\bibinfo{year}{2015}).

\bibitem[{\citenamefont{Luitz et~al.}(2016)\citenamefont{Luitz, Laflorencie,
  and Alet}}]{Luitz2016}
\bibinfo{author}{\bibfnamefont{D.~J.} \bibnamefont{Luitz}},
  \bibinfo{author}{\bibfnamefont{N.}~\bibnamefont{Laflorencie}},
  \bibnamefont{and} \bibinfo{author}{\bibfnamefont{F.}~\bibnamefont{Alet}},
  \bibinfo{journal}{Phys. Rev. B} \textbf{\bibinfo{volume}{93}},
  \bibinfo{pages}{060201} (\bibinfo{year}{2016}).

\bibitem[{\citenamefont{Torres-Herrera and Santos}(2015)}]{Torres2015}
\bibinfo{author}{\bibfnamefont{E.~J.} \bibnamefont{Torres-Herrera}}
  \bibnamefont{and} \bibinfo{author}{\bibfnamefont{L.~F.}
  \bibnamefont{Santos}}, \bibinfo{journal}{Phys. Rev. B}
  \textbf{\bibinfo{volume}{92}}, \bibinfo{pages}{014208}
  (\bibinfo{year}{2015}).

\bibitem[{\citenamefont{Torres-Herrera
  et~al.}(2016)\citenamefont{Torres-Herrera, T\'avora, and
  Santos}}]{Torres2016BJP}
\bibinfo{author}{\bibfnamefont{E.~J.} \bibnamefont{Torres-Herrera}},
  \bibinfo{author}{\bibfnamefont{M.}~\bibnamefont{T\'avora}}, \bibnamefont{and}
  \bibinfo{author}{\bibfnamefont{L.~F.} \bibnamefont{Santos}},
  \bibinfo{journal}{Braz. J. Phys.} \textbf{\bibinfo{volume}{46}},
  \bibinfo{pages}{239} (\bibinfo{year}{2016}).

\bibitem[{\citenamefont{Altman and Vosk}(2015)}]{Altman2015}
\bibinfo{author}{\bibfnamefont{E.}~\bibnamefont{Altman}} \bibnamefont{and}
  \bibinfo{author}{\bibfnamefont{R.}~\bibnamefont{Vosk}},
  \bibinfo{journal}{Ann. Rev. Cond. Mat. Phys.} \textbf{\bibinfo{volume}{6}},
  \bibinfo{pages}{383} (\bibinfo{year}{2015}).

\bibitem[{\citenamefont{Nandkishore and Huse}(2015)}]{Nandkishore2015}
\bibinfo{author}{\bibfnamefont{R.}~\bibnamefont{Nandkishore}} \bibnamefont{and}
  \bibinfo{author}{\bibfnamefont{D.}~\bibnamefont{Huse}},
  \bibinfo{journal}{Annu. Rev. Cond. Mat. Phys.} \textbf{\bibinfo{volume}{6}},
  \bibinfo{pages}{15} (\bibinfo{year}{2015}).

\bibitem[{\citenamefont{Serbyn et~al.}(2015)\citenamefont{Serbyn,
  Papi\ifmmode~\acute{c}\else \'{c}\fi{}, and Abanin}}]{Serbyn2015}
\bibinfo{author}{\bibfnamefont{M.}~\bibnamefont{Serbyn}},
  \bibinfo{author}{\bibfnamefont{Z.}~\bibnamefont{Papi\ifmmode~\acute{c}\else
  \'{c}\fi{}}}, \bibnamefont{and} \bibinfo{author}{\bibfnamefont{D.~A.}
  \bibnamefont{Abanin}}, \bibinfo{journal}{Phys. Rev. X}
  \textbf{\bibinfo{volume}{5}}, \bibinfo{pages}{041047} (\bibinfo{year}{2015}).

\bibitem[{\citenamefont{Singh et~al.}(2016)\citenamefont{Singh, Bardarson, and
  Pollmann}}]{Singh2016}
\bibinfo{author}{\bibfnamefont{R.}~\bibnamefont{Singh}},
  \bibinfo{author}{\bibfnamefont{J.~H.} \bibnamefont{Bardarson}},
  \bibnamefont{and} \bibinfo{author}{\bibfnamefont{F.}~\bibnamefont{Pollmann}},
  \bibinfo{journal}{New J. Phys.} \textbf{\bibinfo{volume}{18}},
  \bibinfo{pages}{023046} (\bibinfo{year}{2016}).

\bibitem[{\citenamefont{Agarwal et~al.}(2015)\citenamefont{Agarwal,
  Gopalakrishnan, Knap, M\"uller, and Demler}}]{Agarwal2015}
\bibinfo{author}{\bibfnamefont{K.}~\bibnamefont{Agarwal}},
  \bibinfo{author}{\bibfnamefont{S.}~\bibnamefont{Gopalakrishnan}},
  \bibinfo{author}{\bibfnamefont{M.}~\bibnamefont{Knap}},
  \bibinfo{author}{\bibfnamefont{M.}~\bibnamefont{M\"uller}}, \bibnamefont{and}
  \bibinfo{author}{\bibfnamefont{E.}~\bibnamefont{Demler}},
  \bibinfo{journal}{Phys. Rev. Lett.} \textbf{\bibinfo{volume}{114}},
  \bibinfo{pages}{160401} (\bibinfo{year}{2015}).

\bibitem[{\citenamefont{Bertrand and
  Garc\'{\i}a-Garc\'{\i}a}(2016)}]{Bertrand2016}
\bibinfo{author}{\bibfnamefont{C.~L.} \bibnamefont{Bertrand}} \bibnamefont{and}
  \bibinfo{author}{\bibfnamefont{A.~M.} \bibnamefont{Garc\'{\i}a-Garc\'{\i}a}},
  \bibinfo{journal}{Phys. Rev. B} \textbf{\bibinfo{volume}{94}},
  \bibinfo{pages}{144201} (\bibinfo{year}{2016}).

\bibitem[{\citenamefont{Gopalakrishnan
  et~al.}(2016)\citenamefont{Gopalakrishnan, Agarwal, Demler, Huse, and
  Knap}}]{Gopalakrishnan2016}
\bibinfo{author}{\bibfnamefont{S.}~\bibnamefont{Gopalakrishnan}},
  \bibinfo{author}{\bibfnamefont{K.}~\bibnamefont{Agarwal}},
  \bibinfo{author}{\bibfnamefont{E.~A.} \bibnamefont{Demler}},
  \bibinfo{author}{\bibfnamefont{D.~A.} \bibnamefont{Huse}}, \bibnamefont{and}
  \bibinfo{author}{\bibfnamefont{M.}~\bibnamefont{Knap}},
  \bibinfo{journal}{Phys. Rev. B} \textbf{\bibinfo{volume}{93}},
  \bibinfo{pages}{134206} (\bibinfo{year}{2016}).

\bibitem[{\citenamefont{Monthus}(2016)}]{Monthus2016}
\bibinfo{author}{\bibfnamefont{C.}~\bibnamefont{Monthus}},
  \bibinfo{journal}{Entropy} \textbf{\bibinfo{volume}{18}},
  \bibinfo{pages}{122} (\bibinfo{year}{2016}), ISSN \bibinfo{issn}{1099-4300}.

\bibitem[{\citenamefont{Gogolin and Eisert}(2016)}]{Gogolin2016}
\bibinfo{author}{\bibfnamefont{C.}~\bibnamefont{Gogolin}} \bibnamefont{and}
  \bibinfo{author}{\bibfnamefont{J.}~\bibnamefont{Eisert}},
  \bibinfo{journal}{Rep. Prog. Phys.} \textbf{\bibinfo{volume}{79}},
  \bibinfo{pages}{056001} (\bibinfo{year}{2016}).

\bibitem[{\citenamefont{Bari\ifmmode \check{s}\else
  \v{s}\fi{}i\ifmmode~\acute{c}\else \'{c}\fi{}
  et~al.}(2016)\citenamefont{Bari\ifmmode \check{s}\else
  \v{s}\fi{}i\ifmmode~\acute{c}\else \'{c}\fi{}, Kokalj, Balog, and
  Prelov\ifmmode~\check{s}\else \v{s}\fi{}ek}}]{Barisic2016}
\bibinfo{author}{\bibfnamefont{O.~S.} \bibnamefont{Bari\ifmmode \check{s}\else
  \v{s}\fi{}i\ifmmode~\acute{c}\else \'{c}\fi{}}},
  \bibinfo{author}{\bibfnamefont{J.}~\bibnamefont{Kokalj}},
  \bibinfo{author}{\bibfnamefont{I.}~\bibnamefont{Balog}}, \bibnamefont{and}
  \bibinfo{author}{\bibfnamefont{P.}~\bibnamefont{Prelov\ifmmode~\check{s}\else
  \v{s}\fi{}ek}}, \bibinfo{journal}{Phys. Rev. B}
  \textbf{\bibinfo{volume}{94}}, \bibinfo{pages}{045126}
  (\bibinfo{year}{2016}).

\bibitem[{\citenamefont{Khemani et~al.}(2016)\citenamefont{Khemani, Pollmann,
  and Sondhi}}]{Khemani2016}
\bibinfo{author}{\bibfnamefont{V.}~\bibnamefont{Khemani}},
  \bibinfo{author}{\bibfnamefont{F.}~\bibnamefont{Pollmann}}, \bibnamefont{and}
  \bibinfo{author}{\bibfnamefont{S.~L.} \bibnamefont{Sondhi}},
  \bibinfo{journal}{Phys. Rev. Lett.} \textbf{\bibinfo{volume}{116}},
  \bibinfo{pages}{247204} (\bibinfo{year}{2016}).

\bibitem[{\citenamefont{Enss et~al.}(2017)\citenamefont{Enss, Andraschko, and
  Sirker}}]{Enss2017}
\bibinfo{author}{\bibfnamefont{T.}~\bibnamefont{Enss}},
  \bibinfo{author}{\bibfnamefont{F.}~\bibnamefont{Andraschko}},
  \bibnamefont{and} \bibinfo{author}{\bibfnamefont{J.}~\bibnamefont{Sirker}},
  \bibinfo{journal}{Phys. Rev. B} \textbf{\bibinfo{volume}{95}},
  \bibinfo{pages}{045121} (\bibinfo{year}{2017}).

\bibitem[{\citenamefont{Torres-Herrera and Santos}(2017)}]{Torres2017}
\bibinfo{author}{\bibfnamefont{E.~J.} \bibnamefont{Torres-Herrera}}
  \bibnamefont{and} \bibinfo{author}{\bibfnamefont{L.~F.}
  \bibnamefont{Santos}}, \bibinfo{journal}{Ann. Phys. (Berlin)} p.
  \bibinfo{pages}{1600284} (\bibinfo{year}{2017}), ISSN
  \bibinfo{issn}{1521-3889}.

\bibitem[{\citenamefont{Torres-Herrera and Santos}()}]{TorresARXIV}
\bibinfo{author}{\bibfnamefont{E.~J.} \bibnamefont{Torres-Herrera}}
  \bibnamefont{and} \bibinfo{author}{\bibfnamefont{L.~F.}
  \bibnamefont{Santos}}, \bibinfo{note}{arXiv:1702.04363}.

\bibitem[{\citenamefont{Luitz and Lev}()}]{LuitzARXIV}
\bibinfo{author}{\bibfnamefont{D.~J.} \bibnamefont{Luitz}} \bibnamefont{and}
  \bibinfo{author}{\bibfnamefont{Y.~B.} \bibnamefont{Lev}},
  \bibinfo{note}{arXiv:1607.01012}.

\bibitem[{\citenamefont{Schreiber et~al.}(2015)\citenamefont{Schreiber,
  Hodgman, Bordia, L\"uschen, Fischer, Vosk, Altman, Schneider, and
  Bloch}}]{Schreiber2015}
\bibinfo{author}{\bibfnamefont{M.}~\bibnamefont{Schreiber}},
  \bibinfo{author}{\bibfnamefont{S.~S.} \bibnamefont{Hodgman}},
  \bibinfo{author}{\bibfnamefont{P.}~\bibnamefont{Bordia}},
  \bibinfo{author}{\bibfnamefont{H.~P.} \bibnamefont{L\"uschen}},
  \bibinfo{author}{\bibfnamefont{M.~H.} \bibnamefont{Fischer}},
  \bibinfo{author}{\bibfnamefont{R.}~\bibnamefont{Vosk}},
  \bibinfo{author}{\bibfnamefont{E.}~\bibnamefont{Altman}},
  \bibinfo{author}{\bibfnamefont{U.}~\bibnamefont{Schneider}},
  \bibnamefont{and} \bibinfo{author}{\bibfnamefont{I.}~\bibnamefont{Bloch}},
  \bibinfo{journal}{Science} \textbf{\bibinfo{volume}{349}},
  \bibinfo{pages}{842} (\bibinfo{year}{2015}).

\bibitem[{\citenamefont{Kondov et~al.}(2015)\citenamefont{Kondov, McGehee, Xu,
  and DeMarco}}]{Kondov2015}
\bibinfo{author}{\bibfnamefont{S.~S.} \bibnamefont{Kondov}},
  \bibinfo{author}{\bibfnamefont{W.~R.} \bibnamefont{McGehee}},
  \bibinfo{author}{\bibfnamefont{W.}~\bibnamefont{Xu}}, \bibnamefont{and}
  \bibinfo{author}{\bibfnamefont{B.}~\bibnamefont{DeMarco}},
  \bibinfo{journal}{Phys. Rev. Lett.} \textbf{\bibinfo{volume}{114}},
  \bibinfo{pages}{083002} (\bibinfo{year}{2015}).

\bibitem[{\citenamefont{Meyer and Wallach}(2002)}]{Meyer:02}
\bibinfo{author}{\bibfnamefont{D.~A.} \bibnamefont{Meyer}} \bibnamefont{and}
  \bibinfo{author}{\bibfnamefont{N.~R.} \bibnamefont{Wallach}},
  \bibinfo{journal}{J. Math. Phys.} \textbf{\bibinfo{volume}{43}},
  \bibinfo{pages}{4273} (\bibinfo{year}{2002}).

\bibitem[{\citenamefont{Brennen}(2003)}]{Brennen:03}
\bibinfo{author}{\bibfnamefont{G.}~\bibnamefont{Brennen}},
  \bibinfo{journal}{Quantum Inf. Comput.} \textbf{\bibinfo{volume}{3}},
  \bibinfo{pages}{619} (\bibinfo{year}{2003}).

\bibitem[{\citenamefont{Paula et~al.}(2013{\natexlab{a}})\citenamefont{Paula,
  de~Oliveira, and Sarandy}}]{Paula:13-1}
\bibinfo{author}{\bibfnamefont{F.~M.} \bibnamefont{Paula}},
  \bibinfo{author}{\bibfnamefont{T.~R.} \bibnamefont{de~Oliveira}},
  \bibnamefont{and} \bibinfo{author}{\bibfnamefont{M.~S.}
  \bibnamefont{Sarandy}}, \bibinfo{journal}{Phys. Rev. A}
  \textbf{\bibinfo{volume}{87}}, \bibinfo{pages}{064101}
  (\bibinfo{year}{2013}{\natexlab{a}}).

\bibitem[{\citenamefont{Nakano et~al.}(2013)\citenamefont{Nakano, Piani, and
  Adesso}}]{Nakano:13}
\bibinfo{author}{\bibfnamefont{T.}~\bibnamefont{Nakano}},
  \bibinfo{author}{\bibfnamefont{M.}~\bibnamefont{Piani}}, \bibnamefont{and}
  \bibinfo{author}{\bibfnamefont{G.}~\bibnamefont{Adesso}},
  \bibinfo{journal}{Phys. Rev. A} \textbf{\bibinfo{volume}{88}},
  \bibinfo{pages}{012117} (\bibinfo{year}{2013}).

\bibitem[{\citenamefont{Paula et~al.}(2013{\natexlab{b}})\citenamefont{Paula,
  Montealegre, Saguia, de~Oliveira, and Sarandy}}]{Paula:13-3}
\bibinfo{author}{\bibfnamefont{F.~M.} \bibnamefont{Paula}},
  \bibinfo{author}{\bibfnamefont{J.~D.} \bibnamefont{Montealegre}},
  \bibinfo{author}{\bibfnamefont{A.}~\bibnamefont{Saguia}},
  \bibinfo{author}{\bibfnamefont{T.~R.} \bibnamefont{de~Oliveira}},
  \bibnamefont{and} \bibinfo{author}{\bibfnamefont{M.~S.}
  \bibnamefont{Sarandy}}, \bibinfo{journal}{EPL (Europhys. Lett.)}
  \textbf{\bibinfo{volume}{103}}, \bibinfo{pages}{50008}
  (\bibinfo{year}{2013}{\natexlab{b}}).

\bibitem[{\citenamefont{Walborn et~al.}(2006)\citenamefont{Walborn,
  Souto~Ribeiro, Davidovich, Mintert, and Buchleitner}}]{Walborn:06}
\bibinfo{author}{\bibfnamefont{S.~P.} \bibnamefont{Walborn}},
  \bibinfo{author}{\bibfnamefont{P.~H.} \bibnamefont{Souto~Ribeiro}},
  \bibinfo{author}{\bibfnamefont{L.}~\bibnamefont{Davidovich}},
  \bibinfo{author}{\bibfnamefont{F.}~\bibnamefont{Mintert}}, \bibnamefont{and}
  \bibinfo{author}{\bibfnamefont{A.}~\bibnamefont{Buchleitner}},
  \bibinfo{journal}{Nature} \textbf{\bibinfo{volume}{440}},
  \bibinfo{pages}{1022} (\bibinfo{year}{2006}).

\bibitem[{\citenamefont{Jurcevic et~al.}(2014)\citenamefont{Jurcevic, Lanyon,
  Hauke, P., Zoller, Blatt, and Roos}}]{Jurcevic:14}
\bibinfo{author}{\bibfnamefont{P.}~\bibnamefont{Jurcevic}},
  \bibinfo{author}{\bibfnamefont{B.~P.} \bibnamefont{Lanyon}},
  \bibinfo{author}{\bibnamefont{Hauke}},
  \bibinfo{author}{\bibfnamefont{C.}~\bibnamefont{P.}, \bibfnamefont{Hempel}},
  \bibinfo{author}{\bibfnamefont{P.}~\bibnamefont{Zoller}},
  \bibinfo{author}{\bibfnamefont{R.}~\bibnamefont{Blatt}}, \bibnamefont{and}
  \bibinfo{author}{\bibfnamefont{C.~F.} \bibnamefont{Roos}},
  \bibinfo{journal}{Nature} \textbf{\bibinfo{volume}{511}},
  \bibinfo{pages}{202} (\bibinfo{year}{2014}).

\bibitem[{\citenamefont{Fukuhara et~al.}(2015)\citenamefont{Fukuhara, Hild,
  Zeiher, Schau\ss{}, Bloch, Endres, and Gross}}]{Fukuhara:15}
\bibinfo{author}{\bibfnamefont{T.}~\bibnamefont{Fukuhara}},
  \bibinfo{author}{\bibfnamefont{S.}~\bibnamefont{Hild}},
  \bibinfo{author}{\bibfnamefont{J.}~\bibnamefont{Zeiher}},
  \bibinfo{author}{\bibfnamefont{P.}~\bibnamefont{Schau\ss{}}},
  \bibinfo{author}{\bibfnamefont{I.}~\bibnamefont{Bloch}},
  \bibinfo{author}{\bibfnamefont{M.}~\bibnamefont{Endres}}, \bibnamefont{and}
  \bibinfo{author}{\bibfnamefont{C.}~\bibnamefont{Gross}},
  \bibinfo{journal}{Phys. Rev. Lett.} \textbf{\bibinfo{volume}{115}},
  \bibinfo{pages}{035302} (\bibinfo{year}{2015}).

\bibitem[{\citenamefont{Ollivier and Zurek}(2001)}]{Ollivier:01}
\bibinfo{author}{\bibfnamefont{H.}~\bibnamefont{Ollivier}} \bibnamefont{and}
  \bibinfo{author}{\bibfnamefont{W.~H.} \bibnamefont{Zurek}},
  \bibinfo{journal}{Phys. Rev. Lett.} \textbf{\bibinfo{volume}{88}},
  \bibinfo{pages}{017901} (\bibinfo{year}{2001}).

\bibitem[{\citenamefont{Henderson and Vedral}(2001)}]{Henderson2001}
\bibinfo{author}{\bibfnamefont{L.}~\bibnamefont{Henderson}} \bibnamefont{and}
  \bibinfo{author}{\bibfnamefont{V.}~\bibnamefont{Vedral}},
  \bibinfo{journal}{J. Phys. A} \textbf{\bibinfo{volume}{34}},
  \bibinfo{pages}{6899} (\bibinfo{year}{2001}).

\bibitem[{\citenamefont{Luo}(2008)}]{Luo:08}
\bibinfo{author}{\bibfnamefont{S.}~\bibnamefont{Luo}}, \bibinfo{journal}{Phys.
  Rev. A} \textbf{\bibinfo{volume}{77}}, \bibinfo{pages}{022301}
  (\bibinfo{year}{2008}).

\bibitem[{\citenamefont{Saguia et~al.}(2011)\citenamefont{Saguia, Rulli,
  de~Oliveira, and Sarandy}}]{Saguia:11}
\bibinfo{author}{\bibfnamefont{A.}~\bibnamefont{Saguia}},
  \bibinfo{author}{\bibfnamefont{C.~C.} \bibnamefont{Rulli}},
  \bibinfo{author}{\bibfnamefont{T.~R.} \bibnamefont{de~Oliveira}},
  \bibnamefont{and} \bibinfo{author}{\bibfnamefont{M.~S.}
  \bibnamefont{Sarandy}}, \bibinfo{journal}{Phys. Rev. A}
  \textbf{\bibinfo{volume}{84}}, \bibinfo{pages}{042123}
  (\bibinfo{year}{2011}).

\bibitem[{\citenamefont{Werlang et~al.}(2010)\citenamefont{Werlang, Trippe,
  Ribeiro, and Rigolin}}]{Werlang:10}
\bibinfo{author}{\bibfnamefont{T.}~\bibnamefont{Werlang}},
  \bibinfo{author}{\bibfnamefont{C.}~\bibnamefont{Trippe}},
  \bibinfo{author}{\bibfnamefont{G.~A.~P.} \bibnamefont{Ribeiro}},
  \bibnamefont{and} \bibinfo{author}{\bibfnamefont{G.}~\bibnamefont{Rigolin}},
  \bibinfo{journal}{Phys. Rev. Lett.} \textbf{\bibinfo{volume}{105}},
  \bibinfo{pages}{095702} (\bibinfo{year}{2010}).

\bibitem[{\citenamefont{Maziero et~al.}(2012)\citenamefont{Maziero, C\'eleri,
  Serra, and Sarandy}}]{Maziero:10}
\bibinfo{author}{\bibfnamefont{J.}~\bibnamefont{Maziero}},
  \bibinfo{author}{\bibfnamefont{L.}~\bibnamefont{C\'eleri}},
  \bibinfo{author}{\bibfnamefont{R.}~\bibnamefont{Serra}}, \bibnamefont{and}
  \bibinfo{author}{\bibfnamefont{M.}~\bibnamefont{Sarandy}},
  \bibinfo{journal}{Phys. Lett. A} \textbf{\bibinfo{volume}{376}},
  \bibinfo{pages}{1540 } (\bibinfo{year}{2012}), ISSN
  \bibinfo{issn}{0375-9601}.

\bibitem[{\citenamefont{Giraud et~al.}(2007)\citenamefont{Giraud, Martin, and
  Georgeot}}]{Giraud:07}
\bibinfo{author}{\bibfnamefont{O.}~\bibnamefont{Giraud}},
  \bibinfo{author}{\bibfnamefont{J.}~\bibnamefont{Martin}}, \bibnamefont{and}
  \bibinfo{author}{\bibfnamefont{B.}~\bibnamefont{Georgeot}},
  \bibinfo{journal}{Phys. Rev. A} \textbf{\bibinfo{volume}{76}},
  \bibinfo{pages}{042333} (\bibinfo{year}{2007}).

\bibitem[{\citenamefont{Modi et~al.}(2012)\citenamefont{Modi, Brodutch, Cable,
  Paterek, and Vedral}}]{Modi:12}
\bibinfo{author}{\bibfnamefont{K.}~\bibnamefont{Modi}},
  \bibinfo{author}{\bibfnamefont{A.}~\bibnamefont{Brodutch}},
  \bibinfo{author}{\bibfnamefont{H.}~\bibnamefont{Cable}},
  \bibinfo{author}{\bibfnamefont{T.}~\bibnamefont{Paterek}}, \bibnamefont{and}
  \bibinfo{author}{\bibfnamefont{V.}~\bibnamefont{Vedral}},
  \bibinfo{journal}{Rev. Mod. Phys.} \textbf{\bibinfo{volume}{84}},
  \bibinfo{pages}{1655} (\bibinfo{year}{2012}).

\bibitem[{\citenamefont{Brodutch and Modi}(2012)}]{Brodutch:12}
\bibinfo{author}{\bibfnamefont{A.}~\bibnamefont{Brodutch}} \bibnamefont{and}
  \bibinfo{author}{\bibfnamefont{K.}~\bibnamefont{Modi}},
  \bibinfo{journal}{Quantum Inf. Comput.} \textbf{\bibinfo{volume}{12}},
  \bibinfo{pages}{0721} (\bibinfo{year}{2012}).

\bibitem[{\citenamefont{Paula et~al.}(2014)\citenamefont{Paula, Saguia,
  de~Oliveira, and Sarandy}}]{Paula:14}
\bibinfo{author}{\bibfnamefont{F.~M.} \bibnamefont{Paula}},
  \bibinfo{author}{\bibfnamefont{A.}~\bibnamefont{Saguia}},
  \bibinfo{author}{\bibfnamefont{T.~R.} \bibnamefont{de~Oliveira}},
  \bibnamefont{and} \bibinfo{author}{\bibfnamefont{M.~S.}
  \bibnamefont{Sarandy}}, \bibinfo{journal}{EPL (Europhys. Lett.)}
  \textbf{\bibinfo{volume}{108}}, \bibinfo{pages}{10003}
  (\bibinfo{year}{2014}).

\bibitem[{\citenamefont{Ciccarello et~al.}(2014)\citenamefont{Ciccarello,
  Tufarelli, and Giovannetti}}]{Ciccarello:14}
\bibinfo{author}{\bibfnamefont{F.}~\bibnamefont{Ciccarello}},
  \bibinfo{author}{\bibfnamefont{T.}~\bibnamefont{Tufarelli}},
  \bibnamefont{and}
  \bibinfo{author}{\bibfnamefont{V.}~\bibnamefont{Giovannetti}},
  \bibinfo{journal}{New J. Phys.} \textbf{\bibinfo{volume}{16}},
  \bibinfo{pages}{013038} (\bibinfo{year}{2014}).

\bibitem[{\citenamefont{Obando et~al.}(2015)\citenamefont{Obando, Paula, and
  Sarandy}}]{Obando:15}
\bibinfo{author}{\bibfnamefont{P.~C.} \bibnamefont{Obando}},
  \bibinfo{author}{\bibfnamefont{F.~M.} \bibnamefont{Paula}}, \bibnamefont{and}
  \bibinfo{author}{\bibfnamefont{M.~S.} \bibnamefont{Sarandy}},
  \bibinfo{journal}{Phys. Rev. A} \textbf{\bibinfo{volume}{92}},
  \bibinfo{pages}{032307} (\bibinfo{year}{2015}).

\end{thebibliography}

\end{document}